# The design of the wide field monitor for LOFT

S. Brandt[*a], M. Hernanz[b], L. Alvarez[b], A. Argan[e,f], B. Artigues[b], P. Azzarello[c], D. Barret[d], E. Bozzo[c], Budtz-Jørgensen[a], R. Campana[e,f], , A. Cros[d], E. del Monte[e,f], I. Donnarumma[e,f], Y. Evangelista[e,f], M. Feroci[e,f], J.L. Galvez Sanchez[b], D. Götz[g], F. Hansen[a], J.W. den Herder[h], R. Hudec[i], J. Huovelin[j], D. Karelin[b], S. Korpela[j], N. Lund[a], M. Michalska[k], P. Olsen[a], P. Orleanski[k], S. Pedersen[a], M. Pohl[l], A. Rachevski[o], A. Santangelo[m], S. Schanne[g], C. Schmid[n], S. Suchy[m], C. Tenzer[m], A. Vacchi[o], D. Walton[q], J. Wilms[n], G. Zampa[o], N. Zampa[o], J. in't Zand[h], S. Zane[q], A. Zdziarski[p], F. Zwart [h]

[a]DTU Space, Elektrovej Building 327, DK-2800, Kgs. Lyngby, Denmark;
[b]IEEC/CSIC, Campus UAB, E-08193, Bellaterra, Spain;
[c]ISDC, University of Geneva, Chemin d'Ecogia 16, CH-1290, Versoix, Switzerland;
[d]IRAP, 9, Avenue du Colonel Roche, BP 44346, 31028 Toulouse Cedex 4, France;
[e]INAF/IAPS, Via Fosso del Cavaliere 100, I-00133, Roma, Italy;
[f] INFN/Sezione di Roma 2, Via della Ricerca ScientiScientica 1, I-00133, Roma, Italy;
[g]CEA Saclay, DSM/DAPNIA/Service d'Astrophysique, 91191 Gif sur Yvette, France;
[h]SRON, Sorbonnelaan 2, 3584 CA Utrecht, The Netherlands;
[i]Astronomical Institute of the ASCR, v.v.i. and Czech Technical Univ. in Prague, Czech Republic;
[j]Department of Physics, University of Helsinki, Gustaf Hällströmin katu 2a, FI-00560, Helsinki;
[k]Space Research Centre, Polish Academy of Sciences, Bartycka 18A, Warsaw, Poland;
[l]DPNC, University of Geneva, Switzerland;
[m]IAAT, Sand 1, D-72076, Tübingen, Germany;
[n]University of Erlangen-Nuremberg & Erlangen Centre of Astroparticle Physics;
[o]INFN-Trieste, Padriciano 99, I-34127, Trieste, Italy;
[P]N. Copernicus Astronomical Center, Bartycka 18, 00-716 Warsaw, Poland;
[q]Mullard Space Science Laboratory, UCL, Holmbury St Mary, Dorking, Surrey, RH56NT, UK;

## ABSTRACT

LOFT (Large Observatory For x-ray Timing) is one of the ESA M3 missions selected within the Cosmic Vision program in 2011 to carry out an assessment phase study and compete for a launch opportunity in 2022-2024. The phase-A studies of all M3 missions were completed at the end of 2013. LOFT is designed to carry on-board two instruments with sensitivity in the 2-50 keV range: a 10 $m^2$ class Large Area Detector (LAD) with a <1° collimated FoV and a wide field monitor (WFM) making use of coded masks and providing an instantaneous coverage of more than 1/3 of the sky. The prime goal of the WFM will be to detect transient sources to be observed by the LAD. However, thanks to its unique combination of a wide field of view (FoV) and energy resolution (better than 500 eV), the WFM will be also an excellent monitoring instrument to study the long term variability of many classes of X-ray sources. The WFM consists of 10 independent and identical coded mask cameras arranged in 5 pairs to provide the desired sky coverage. We provide here an overview of the instrument design, configuration, and capabilities of the LOFT WFM. The compact and modular design of the WFM could easily make the instrument concept adaptable for other missions.

---

[*] sb@space.dtu.dk; phone +45 4525 9710; www.space.dtu.dk

# 1. INTRODUCTION

LOFT (Large Observatory For x-ray Timing) [10][11][13] is a mission designed to study strong gravity in the vicinity of black holes and to unravel the puzzle of the equation of state of dense matter in neutron stars.

LOFT was one of the M3 missions selected in 2011 to carry out an assessment phase study and compete for a launch opportunity in 2022-2024 [1]. The LOFT phase-A study was successfully completed in late 2013 [13], but LOFT was not selected as the mission to be implemented and launched in 2022-2024. The LOFT Consortium has planned to re-propose LOFT for other launch opportunities.

LOFT is designed to carry on-board two science instruments, both making use of innovative large-area Silicon drift detectors (SDDs). The Large Area Detector (LAD) is a collimated, narrow FoV instrument with an effective area of ~10m$^2$, optimally designed to perform simultaneous X-ray timing and spectroscopic analyses with unprecedented signal-to-noise ratio [10][11]. Through its innovative technologies, the LAD will provide an effective area at least ~15 times larger than any previous flown X-ray mission. The spectral resolution of the LAD will be ~240 eV or better at end of life, and have an absolute timing accuracy < 2μs.

The second instrument on LOFT is a Wide Field Monitor (WFM). This instrument is based on the coded mask principle and makes use of SDDs similar SDDs to those employed for the LAD detectors. The design of the WFM has been optimized to provide the required combination of wide FoV, sensitivity, and source position determination. The WFM is primarily needed to detect interesting X-ray targets (known sources changing their X-ray emission state or new transient sources) to be observed with the LAD. However, the unique combination of sensitivity and large instantaneous FoV of this instrument make the WFM capable of conducting excellent scientific investigations by in its own right.

The LOFT mission is planned to be operational for 3+2 years. This duration is mainly driven by the expected occurrence of rare bright black hole transient outbursts, which are among the prime targets to be studied with the LAD. The large fraction of the sky that can be monitored with the WFM at once and that is instantaneously accessible by the LAD secure that a sufficient number of these transient sources can be observed within the mission life time.

This paper describes the WFM instrument in the configuration that has been optimized for the LOFT mission throughout its phase-A study. Due to its modular design, the WFM concept might be adapted to serve in other contexts.

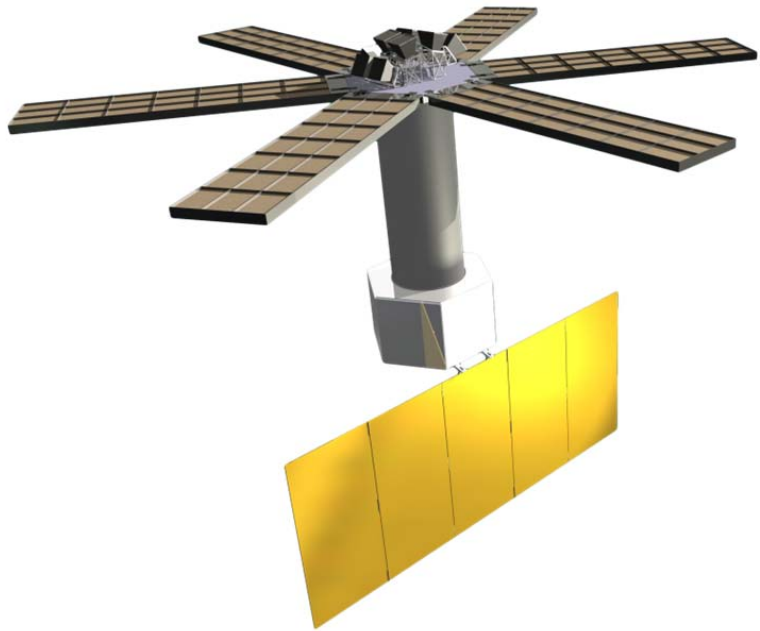

Figure 1 Schematic views of the LOFT spacecraft showing the deployed Large Area Detector (LAD) panels attached to the optical bench. The Wide Field Monitor (WFM) is placed at the center of the optical bench. The viewing direction of the LAD is co-aligned with the direction of maximum response of the WFM.

## 2. SCIENCE OBJECTIVES

The main objectives of the LOFT mission are to study the effects of strong gravity and the properties of ultra-dense matter [13][10]. This will be accomplished by the combined capabilities of the two instruments: the Large Area Detector (LAD) [20] and the Wide Field Monitor (WFM).

### 2.1 LAD Science Objectives

The LAD is designed to: (i) reconstruct the equation of state of ultra-dense matter in neutron stars and (ii) measure the effects of strong gravity in the stationary fields of neutron stars and black holes at time scales relevant to the orbital motion of matter in their closest vicinity.

The neutron star structure and equation of state of ultra-dense matter is studied by measuring the mass and radius of neutron stars with the highest precision achieved so far. This is achieved by using a number of independent techniques.

The conditions of strong-field gravity are explored by measuring the mass and spin of black holes, and by detecting general relativistic precession and the Quasi-Periodic-Oscillations introduced by matter orbiting black holes. The effects of general relativity close to the event horizon of black holes are also studied by observing the distortion of the Fe line, thus demanding a high spectral resolution of the instrument.

Common for these timing studies is the fact that the coherence time of the phenomena is short, which means that an observation with a large area detector cannot be substituted by a longer observation with a smaller detector. Also, many of the phenomena are associated with transient sources or peculiar emission states of known sources, which makes a wide field monitoring instrument mandatory to fulfill the LOFT science goals.

### 2.2 WFM Science Objectives

The LOFT WFM will primarily detect new transient sources and identify emission state changes of known targets that are suitable for observation with the LAD. Thanks to its unique capabilities, the WFM will also be able to carry out important scientific investigations on its own.

With a very large FoV, covering more than 1/3 of the sky at once, the WFM offers a high duty cycle compared to other X-ray monitors with a scanning mode of operation (e.g., the past RXTE/ASM and the current MAXI monitor). The WFM will thus be particularly well suited to identify and localize all those bright impulsive events that are rarely detected by scanning instruments, as Type-I X-ray bursts, Gamma Ray Bursts (GRBs), Soft Gamma Repeater (SGR) flares, and Terrestrial Gamma Flashes (TGFs). The WFM offers the possibility to study these events in a broad energy range (2-50 keV) and with a spectral (timing) resolution down to 300-500 eV (10 μs).

The WFM source location accuracy (<1 arcmin) and energy resolution are well suited to investigate some of the more debated issues related to GRBs. Another area where the WFM will contribute is in the topic of X-ray flashes. WFM studies of the population and properties of X-ray flashes found to accompany supernova shock break-out and the disruption of stars and planetary objects by super-massive black holes is expected to provide crucial results.

With a sensitivity of better than 5 mCrab in one day for observations carried out toward the Galactic Center, the WFM will monitor the long term variability of a large number of X-ray sources (about 100 a day). In addition, the WFM will be endowed with a real time burst alert system, capable of providing the location (~1' accuracy) and trigger time of any impulsive event detected in its FoV within 30s from the discovery to the end user on the ground. This is a unique capability beyond the role of the WFM as a support instrument for the LAD [16].

## 3. LOFT WIDE FIELD MONITOR REQUIREMENTS

The LOFT spacecraft design was studied for ESA by two independent industrial consortia, which came up with somewhat different designs. Both designs, however, planned to have the LAD SDDs spread on deployable panels attached to an optical bench at the top of the spacecraft [11][20], and placed the WFM at the center of the optical bench (see Figure 1 and Figure 2). In these concepts, the FoV of the WFM is limited to the hemisphere centered on the pointing direction of the LAD.

In order to meet the top level science goals of the LOFT mission [10][11][13][20], a set of requirements has been defined for the WFM. These requirements, as well as the performance goals, are summarized in Table 1.

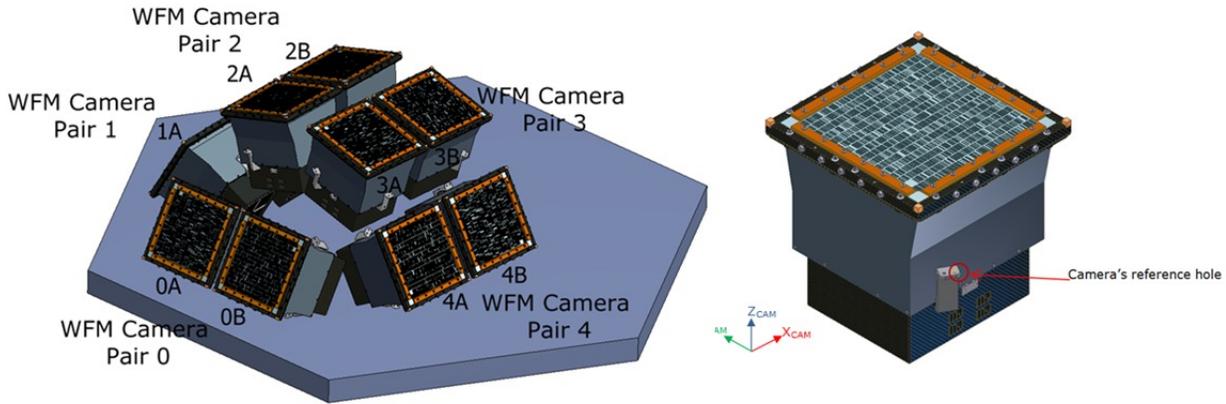

Figure 2 *Left:* the Wide Field Monitor assembly (10 cameras) mounted on the optical bench at the top of the spacecraft. The LAD panels attached to the edge of the bench are not shown. The assembly is protected from direct sun light by a shield (not shown). *Right:* one WFM camera and its physical reference frame.

Table 1 Summary of the WFM requirements and anticipated performances of the studied design

| Item | Requirement | Anticipated performance |
|---|---|---|
| Location accuracy (10 mCrab in 50 ks) | <1 arcmin | <1 arcmin |
| Angular resolution | <5 arcmin | <4.5 arcmin |
| Peak sensitivity in LAD direction (5σ) | 1 Crab (1 s) <br> 5 mCrab (50 ks) | <0.6 Crab (1s) <br> <3 mCrab (50 ks) |
| Absolute flux accuracy | 20 % | <20% |
| Field of View | 1 π steradians around the LAD pointing | 1.75 π steradians at 0% response, 1.33 π steradians at 20% response |
| Energy range | 2 – 50 keV primary | 2 – 50 keV |
| Energy resolution | 500 eV (FWHM) eV @ 6 keV | <300 eV @ 6 keV |
| Energy scale | 4% | <2% |
| Energy bands for compressed images | >=8 | >=8, <=64 |
| Time resolution | 300 s for images <br> 10 μs for event data | <= 300 s for images <br> < 10 μs for event data |
| Absolute time calibration | 2 μs | <2 μs |
| Burst trigger scale | 0.1 s - 100 s | 10 ms - 300 s |
| Rate meter data resolution | 16 ms | <= 16 ms |
| Availability of triggered WFM data | 3 hours | <3 hours |
| On-board memory | 5 min @ 100 Crab | >10 min @ 100 Crab |
| Broadcast of trigger time and position to end user | < 30 s after the event for 65% of the events. Transient position accuracy ~1arcmin | < 25 s after the event for 65% of the events. Transient position accuracy <1arcmin |
| Relative flux calibration precision | 5% | <5% |
| Sensitivity variation knowledge | The WFM sensitivity variation over the mission <10%. | The WFM sensitivity variation over the mission <10%. |
| Number of triggers | Up to 5 GRB triggers per day | >1 GRB triggers per orbit |
| Redundancy | No full loss of FoV | No full loss of FoV |

# 4. WIDE FIELD MONITOR INSTRUMENT OVERVIEW

The WFM is based on the coded mask principle, as employed successfully in different energy bands on several previous missions like, GRANAT/SIGMA, BeppoSAX/WFC, INTEGRAL/JEM-X/IBIS/SPI, RXTE/ASM, SWIFT/BAT, and SuperAGILE. The mask shadow-gram is recorded by a position sensitive detector, and can then be deconvolved into a sky image. The size of the point spread function in the sky image is determined by the ratio of the mask pixel size and the mask-to-detector distance. The mask pixel size must always be larger than the corresponding detector resolution. Previous missions mentioned above include instruments where the detector has a regular 2 dimensional position resolution (for example pixelated detectors), as well as strictly 1-dimensional detectors that can only produce 1-dimensional sky images. In the 1D case, two orthogonally oriented instruments or a rotating instrument are needed to provide accurate source positions in two dimensions.

The WFM is based on a modular design with 10 identical cameras. The WFM detector employs the same basic detector technology as the LAD, but with a read-out scheme that enables 2D position determination of the X-ray interactions in the detector plane required for X-ray imaging. By design, the detector position resolution in one direction is much finer (<60 μm) than in the other direction (<8 mm). This is reflected in the elongated point spread function of the de-convolved sky images. The 10 WFM cameras are grouped into 5 Camera pairs with two co-aligned cameras rotated 90° around the viewing direction with respect to each other. Each camera and pair will cover a ~90°×90° area of the sky at zero response. This combination can achieve the required 1 arcmin source localization accuracy in two dimensions (a margin for the spacecraft pointing errors is included in the budget). The five pairs are oriented to cover a large fraction of the sky accessible to pointed observations with the LAD. The imaging capabilities of the WFM are also needed to resolve any source confusion in the ~1° FoV of the non-imaging LAD instrument. Therefore the WFM will have its peak sensitivity in the LAD pointing direction. The WFM characteristics are summarized in Table 2.

Table 2 Summary of the main WFM characteristics

| WFM Instrument Characteristic | |
|---|---|
| Detector type | Si Drift |
| Mass (10 cameras forming 5 pairs + ICU + harness ) | 125 kg, including 20% margins |
| Peak Power | 109 W |
| Detector Operating Temperature | ÷30 °C < ÷3 °C |
| Total Detector Effective Area (10 cameras) | 1820 cm$^2$ |
| Energy range [keV] | 2-50 keV |
| Energy resolution [FWHM] | <500 eV @ 6 keV |
| Mask pixel size | 250 μm × 16.4 mm |
| FoV | 180° × 90° FWZR plus 90° × 90° towards anti-Sun hemisphere |
| Angular Resolution (geometric) | <5 arcmin (5 arcmin × 5° per camera) |
| Typical/Max data rate after binning and compression | 50/100 kbits/s |

## 4.1 WFM functional design

Each camera is composed of 1 Detector Tray with 4 Silicon Drift Detectors, 4 Beryllium windows, 4 Front-End Electronics, 1 Back-End Electronics assembly, 1 Collimator, and 1 Coded Mask with a Thermal Blanket. The camera pairs and the cameras in the WFM are organized to achieve a high level of redundancy.

Figure 3 shows the functional block diagram for the WFM. The functions of the main components can be summarized as follows:

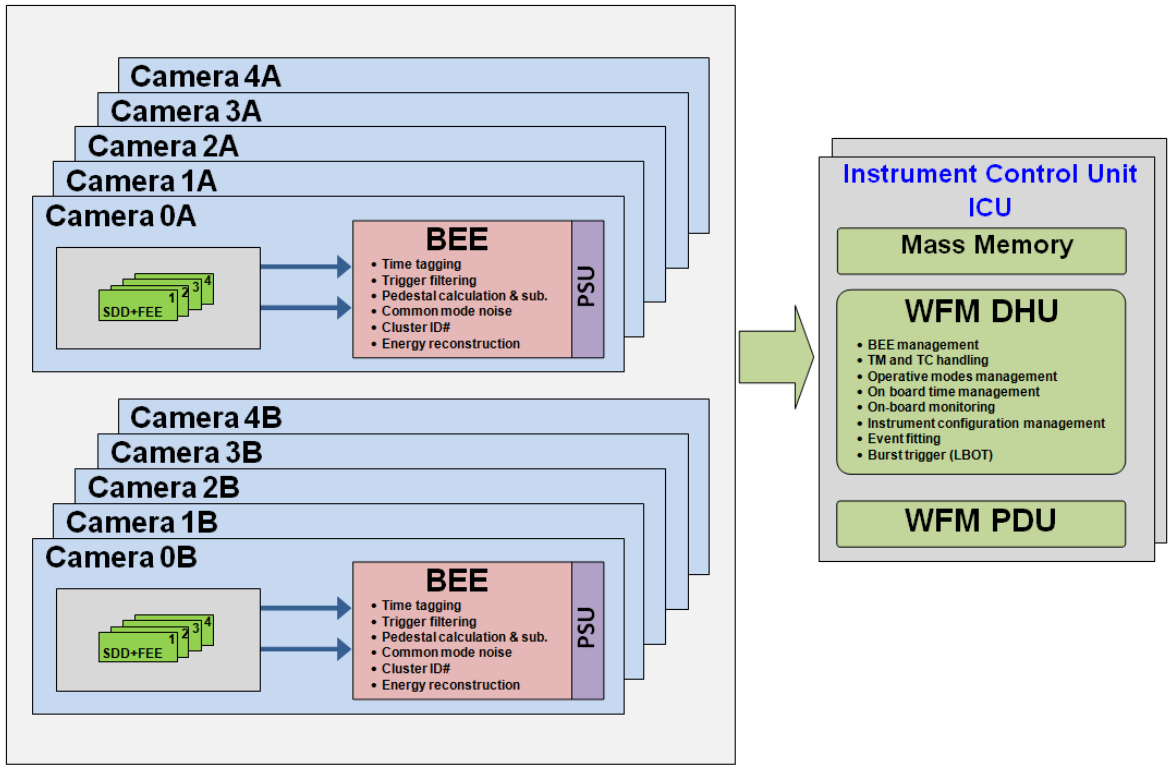

Figure 3 Functional block diagram of the WFM instrument, consisting of 10 identical cameras in 5 pairs and 2 instrument control units including the power distribution unit and mass memory (one ICU in cold redundancy).

The main functions of the WFM FEE, Front End Electronics, are:
- Forward filtered bias voltages to the SDD, Silicon Drift Detector
- Provide power and configuration data to the ASICs
- Read-out and A/D convert the SDD signals (inside ASICs)
- Interface the Back-End Electronics, BEE
- Mechanical support for the SDD

The main functions of the BEE, Back End Electronics are:
- Time tagging of the X-ray events
- Trigger selection
- Pedestal subtraction
- Common mode noise subtraction for the fit of the X-ray signal
- Determination of charge cloud center and width (position in the fine and coarse direction)
- Reconstruction the total charge collected, correcting for channel gains

The main functions of the WFM ICU, Instrument Control Unit, are:
- Interfacing the BEEs
- TC and configuration handling
- On board time management
- Image integration
- Burst triggering
- Power distribution
- Mass memory for storage of telemetry data

## 4.2 Accommodation on the spacecraft

The WFM configuration on LOFT has been optimized to fulfil the support function for the LAD observations. Primarily for thermal reasons, the LAD targets must be chosen within a band on the sky perpendicular to the instantaneous Sun vector. Four of the five WFM camera pairs are therefore set to monitor a 180° arc along this sky band as shown in Figure 2. Covering a larger fraction of the band is not practical due to the sky obscuration caused by the LAD itself. A fifth WFM camera pair has been placed to allow monitoring in the "anti-Sun" direction. This fifth pair will allow monitoring interesting regions on the sky (for example the Galactic Centre) semi-continuously for 9 months each year. In case an interesting source turns on, it will be possible for the LAD to carry out observations of limited duration also in the anti-Sun direction. All 10 cameras are identical in their mechanical, thermal and electrical design. The WFM baseline configuration with partly overlapping fields of view, as illustrated in Figure 2, will cover ~5.5 steradians (~44% of the sky at zero response), and ~4.2 steradians (i.e. 1/3 of the sky) at 20% response relative to on-axis.

Four of the five camera pairs are arranged such that their viewing axes lie in the plane of the solar panel of the LOFT spacecraft. The fifth pair is tilted out of this plane away from the Sun by 60˚. The viewing directions of the four pairs are off-set by ±15˚ and ±60˚ relative to the LAD pointing direction, which also lies in the solar panel plane (see Figure 1). The effective area of the full WFM assembly is shown in the right hand panel of Figure 4. With this arrangement, the two central pairs have the LAD target in their field, where the detectors are fully illuminated, providing the maximum WFM coverage (~160 cm$^2$ effective area in the direction observed by the LAD). The zero response FoV of the 4 pairs extends to ±105˚ × ±45˚ (210˚ × 90˚); however, depending on the configuration of the LAD panels and the placement of the WFM pairs on the optical bench, only an unobstructed FoV of 180˚ × 90˚ can be assured. The 60° tilt of the two outer pairs with respect to the LAD pointing direction is preferred in order to have a reasonable response at the edge defined by the plane of the optical bench. In this configuration, the WFM nominally covers half of the sky that is accessible to LAD pointings. The WFM may therefore cover all the sky accessible to the LAD in 2 LOFT pointings (this is equivalent to at least 75% of the sky).

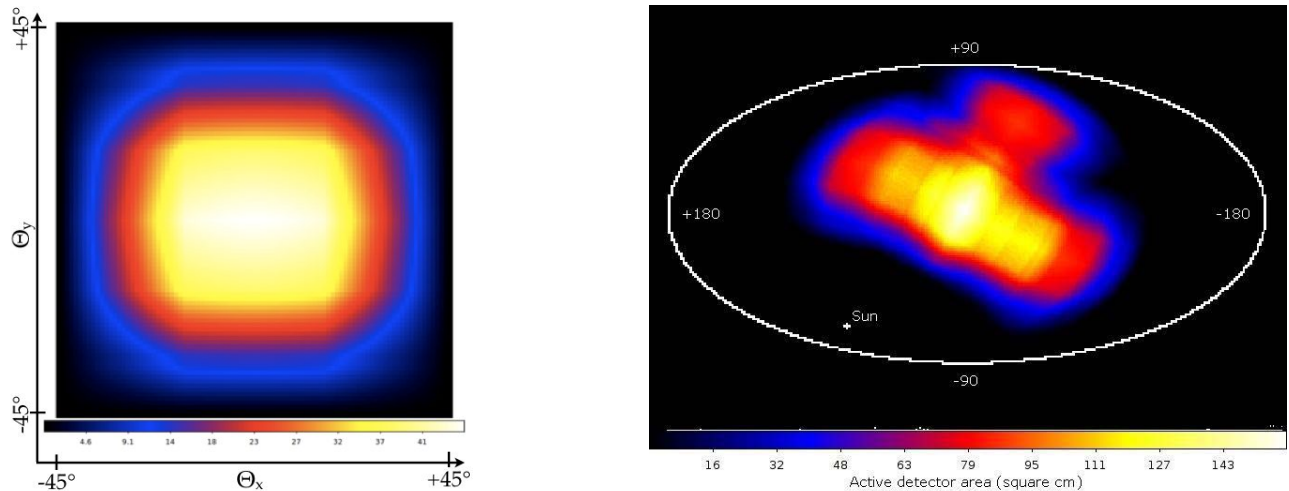

Figure 4 *Left panel:* map of the single camera sensitive area expressed in cm$^2$, with a maximum of ~45 cm$^2$ of a single camera. The map takes into account the main geometrical effects (mask open fraction, vignetting, shadowing of the collimator walls, detector non-sensitive areas, and projection effects). *Right panel:* map in Galactic coordinates of the active detector area for an example observation of the Galactic center. The effective area has its maximum of ~160 cm$^2$ in the direction of the LAD pointing. The figure does not include any obscuration at the edge of the FoV by the optical bench.

We note that the WFM configuration has several overlapping fields of view between different pairs. This will provide a good handle of the in-flight calibration of the off-axis response, as sources will be simultaneously observed at different off-axis angles in individual cameras.

A stable thermal environment is essential for the detector performance, but also for the mechanical stability of the collimator and mask assembly. A sunshade will thus be placed on the optical bench to prevent direct illumination of the WFM masks by the Sun during normal observations.

### 4.3 WFM camera and coded mask design

In a coded mask instrument, photons from a certain direction in the sky project the mask pattern on a position sensitive detector. In focusing optics the signal from a source is concentrated on a small detector area, while for the coded mask the signal is distributed over a large detector area, where the source can illuminate the detector through the mask.

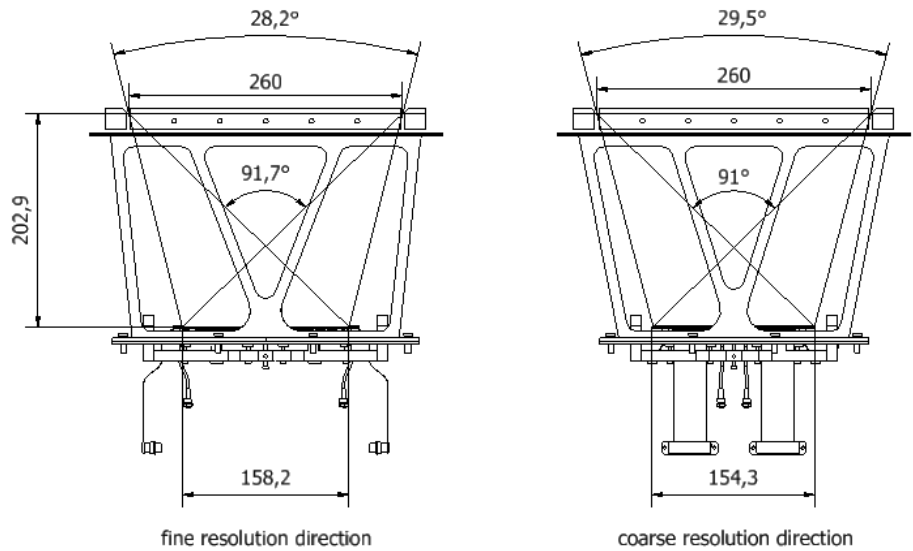

Figure 5 Optical configuration of a WFM camera showing the size of the mask and the detector plane and the distance between them (in mm). The angles for the fully illuminated and zero response angles are also shown.

The structure of an individual WFM camera follows the classical design of coded mask instruments [17][19], as illustrated in Figure 5. However, the position resolution of the detector used in the WFM is very asymmetric, being <60 μm in one direction and <8 mm in the orthogonal direction. This is reflected in the coded mask, which has a pattern consisting of 1040 × 16 open/closed elements with a mask pitch of 250 μm × 16.4 mm. The dimensions of the open elements are 250 μm × 14 mm with 2.4 mm spacing between the elements in the coarse resolution direction for mechanical reasons. The mask nominal open fraction is chosen to be ~25% in order to improve sensitivity to weaker sources and reduce the required telemetry bandwidth. The coded mask consists of a 150 μm thick Tungsten foil. The mask design is illustrated in Figure 6.

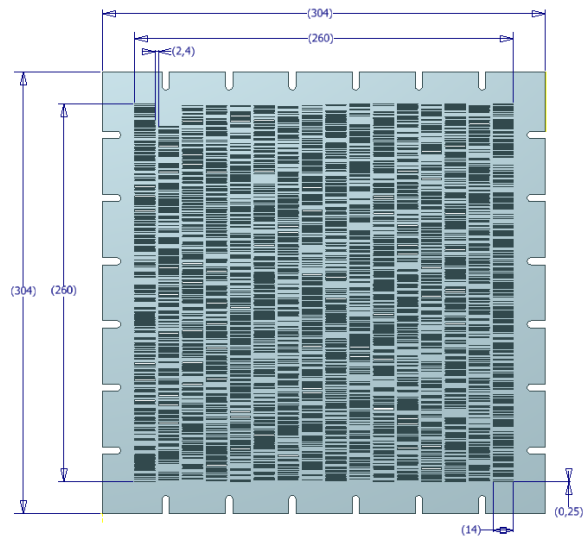

Figure 6 The WFM coded mask design (dimensions are shown in mm)

The detector-mask distance is 202.9 mm. The angular resolution (FWHM) is for the on-axis viewing direction equal to the ratio of the mask pitch and the detector to mask distance. Thus the angular resolution in the direction of the fine mask pitch is 4.24 arcmin and in the coarse direction it is 4.6 degrees. We note that the point source location accuracy scales with the ratio between the signal to noise and the angular resolution. In addition errors in source position determination are introduced by deformation in the camera geometry, introducing a need for a sun shield for the WFM. The mask must be flat with a tolerance of ±50 μm over its entire surface across the full operational temperature range.

Table 3 summarizes the optical design and the performance of a WFM camera, as well as the properties of a WFM camera pair. Extensive simulations of the WFM imaging performance have been carried out [7][8][9]. Figure 7 shows simulated images of a single bright source. The map represents a central region (~10° x 10°) of the camera pair FoV obtained by combining the sky observed simultaneously by the two cameras.

Table 3 Optical design and performance of each WFM camera and camera pair

|  | **Camera** | **Camera Pair** |
|---|---|---|
| **Optical design** | | |
| Mask-detector distance | 202.9 mm | |
| Mask size | $260 \times 260 mm^2$ | |
| Mask pitch | $0.250 \times 16.4 mm^2$ | |
| Size of open Mask elements | $0.250 \times 14 mm^2$ | |
| Active detector area | $182\ cm^2$ | $364\ cm^2$ |
| Peak Effective Area (on-axis, through mask) | $>39\ cm^2$ | $> 78\ cm^2$ |
| SDD spatial resolution (fine direction, FWHM) | < 60 μm | |
| SDD spatial resolution (coarse direction, FWHM) | < 8 mm | |
| **Optical Performance** | | |
| Angular resolution | 4.24' × 4.6° | 4.24' × 4.24' |
| Point source location accuracy (SNR>10 σ) | 0.9' × 30' | 0.9' × 0.9' |
| FoV (detector fully illuminated) | 28.2° × 29.5° | 28.2° × 28.2° |
| FoV (full width at zero response) | 91.7° × 91.0° | 91.7° × 91.7° |

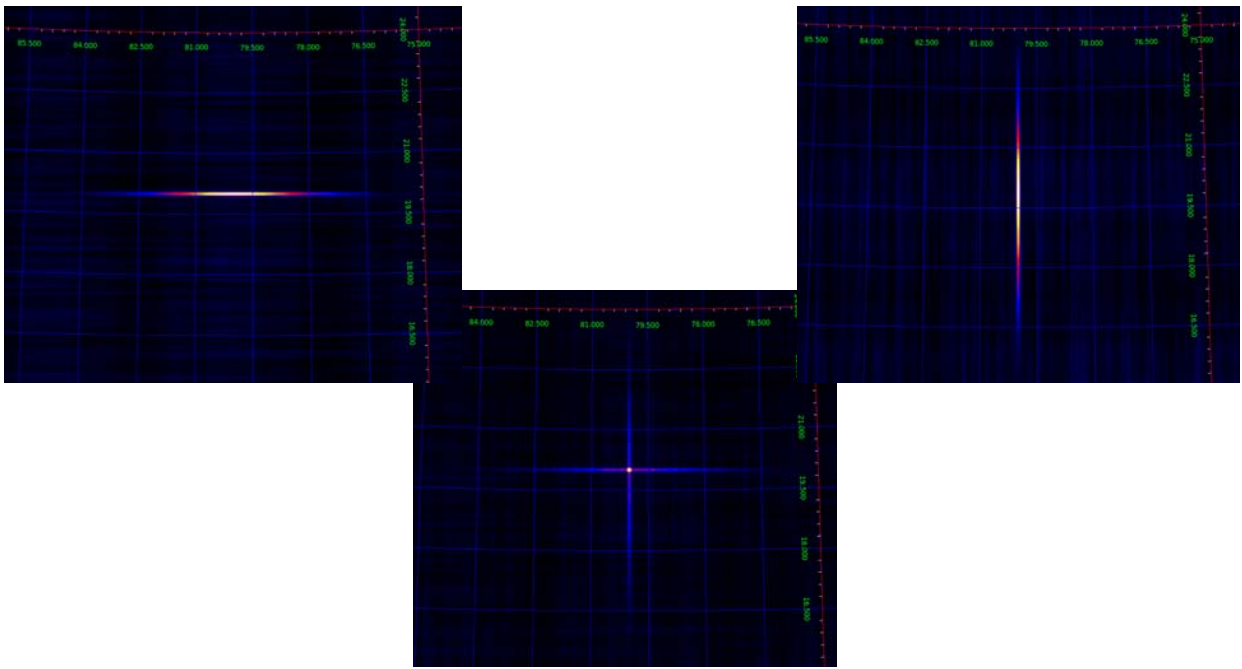

Figure 7 A simulated 2D image of an isolated source obtained by combining the images from two cameras in a pair shown in the upper part. Only a ~10° × 10° part of the FoV is shown.

# 5. THE WFM ELECTRICAL DESIGN

The electrical architecture of the LOFT WFM is shown schematically in Figure 8 and encompasses all electrical subsystems of the WFM and their interfaces with the LOFT spacecraft platform.

Electrically, the central part of the WFM is the instrument control unit (ICU), containing the data handling unit (DHU) with mass memory and a power distribution unit (PDU). The ICU includes a main and a redundant unit housed in two separate boxes as shown in the schematic. The LOFT burst on-board trigger (LBOT) functionality is implemented as a part of the DHU.

The DHU is directly attached to the LOFT Spacecraft on-board data handling (OBDH) system. The electrical interface is assumed to be SpaceWire. The SpaceWire connections are cross-strapped between the main and redundant DHU and the main and redundant OBDH. Mass memory for storing WFM data until downlink is an integral part of the DHU. Each camera back end electronic (BEE) is connected to both the main and redundant DHU, both for control, data, and monitoring. The bus power is routed through the WFM power distribution unit (PDU) providing ON/OFF switching and protection capability.

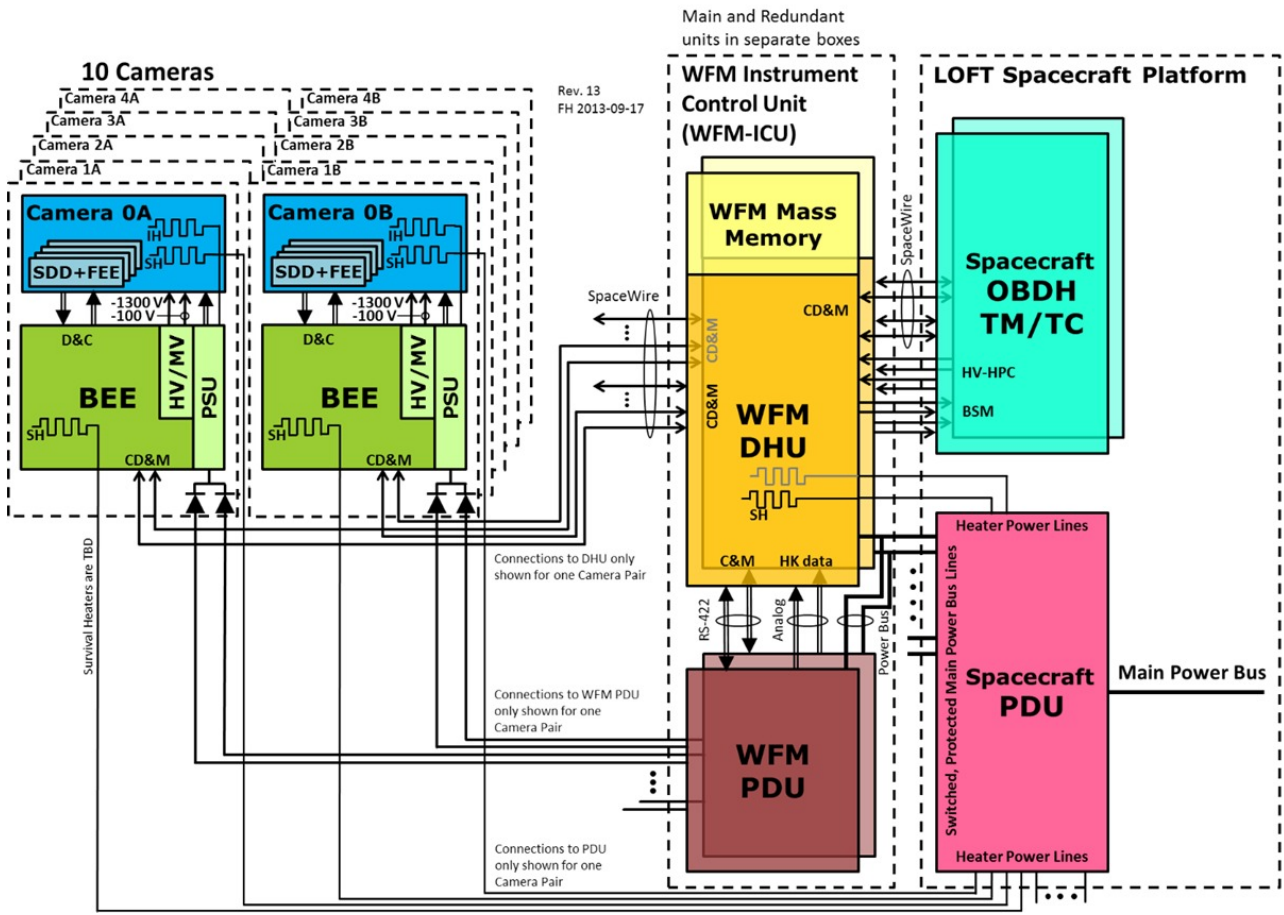

Figure 8 The WFM electrical design with interfaces to the spacecraft onboard data handling unit and power supply unit.

## 5.1 The WFM Silicon Drift Detector

The Silicon Drift Detectors (SDDs) to be used for the WFM have the same design and characteristics as those for the LAD [15], with the only difference in the smaller overall size of the Si tile (for the WFM: 77.4 mm × 72.5 mm, vs 120.8 mm × 72.5 mm for the LAD [20]) and in the smaller anode pitch (145 μm for the WFM, versus 970 μm for the LAD). The LAD and WFM SDDs share the heritage from the ALICE particle experiment in the LHC at CERN [14][18]. The individual SDD for the WFM has the characteristics listed in Table 4. The WFM SDDs are expected to meet the performance goal of a 300 eV spectral resolution.

The key difference between the LAD and the WFM requirements to the detector is the WFM need for position information of the X-ray interactions in order to function as a coded mask instrument. The working principle of the LOFT SDDs is shown in Figure 9.

The size of the charge cloud resulting from an X-ray interaction in the Si depends on the drift length before reaching the anodes. By fitting the charge cloud it is possible to derive three parameters about the incoming photon: (X,Y,E)=(position in the anode direction, drift length, energy). It is possible to achieve a position resolution of ~60 μm (FWHM) in the anode direction and of 3 mm (FWHM) at 6 keV and 8 mm at 2 keV in the drift direction. The imaging performance of the SDD is described in further details in [15] and references therein.

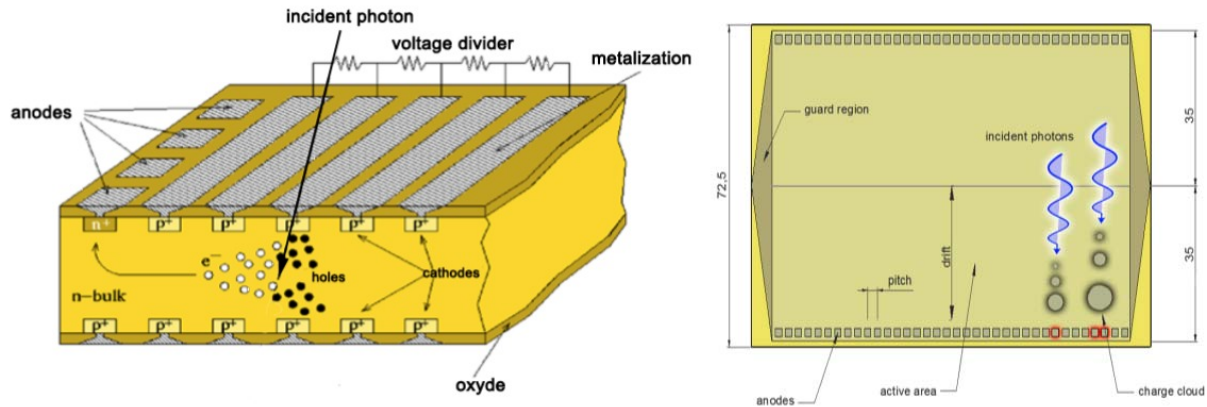

Figure 9 Illustrations of the working principle of the Silicon drift detectors. The absorption of a photon in the Silicon bulk is shown to the left. To the right is shown the detector module with read-out anodes at the top and the bottom. The size of the charge cloud is a function of the drift length and is fitted to provide the position in the drift direction

A total of 4 SDDs compose the detector plane of each individual WFM camera. The overall dimension of the SDD assembly is 145 mm × 154.8 mm. The active area of each WFM camera is a square of 142.5 mm × 142.5 mm. This allows arranging two identical cameras with a 90° relative rotation (in order to achieve fine angular resolution in two coordinates) but still having the same FoV to compose one WFM pair.

The choice of the SDD size is driven by the requirement of a square active area for the overall camera and by the requirement of not increasing the drift length longer than 35 mm. The choice of the anode pitch of 145 m is the result of an optimization study of the detector performance, based on experimental tests and by simulations [10]. The maximum effective area of a detector module is ~45 cm$^2$ and it is shown as function of energy in Figure 10.

Table 4 Summary of the characteristics of one SDD detector module. One WFM camera will contain 4 detector modules

| Parameter | Value |
|---|---|
| Si thickness | 450 μm |
| Si tile geometric size | 77.4 mm × 72.5 mm |
| Si tile active area | 65.1 mm × 70.0 mm = 45.57 cm$^2$ |
| Anode pitch | 145 μm |
| Number of read-out anodes per tile | 448 x 2 rows = 896 total |
| Drift length | 35 mm maximum |
| Anode capacitance | 85 fF |

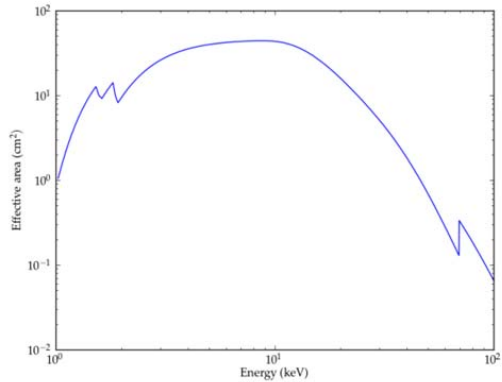

Figure 10 The effective area of one WFM detector module as function of energy. The area reaches ~45 cm$^2$ at 8 keV. As the WFM coded mask has an open fraction of ~25%, this figure also corresponds to the on-axis effective area of one WFM camera which contains 4 detector modules (at least for energies below 30 keV).

### 5.2 Front End Electronics and ASICs

The ASICs on the FEE-board interface directly to the SDD anode points – 448 anode points with a pitch of 145 μm located at two opposite edges of the SDD. Electrically each SDD is divided into two independent halves by the centrally placed high-voltage electrode. The read-out of the 2×448 anode signals in one detector module is performed by 2 rows of ASICs placed on the front end electronics board. The latter is mounted on the back side of the Si detector, as illustrated in Figure 11. The number of channels per ASIC may be 32 or 64, thus requiring 2×14 or 2×7 ASICs. The details of the ASIC design, which to a large extent is common with the LAD, are given in [5].

X-ray photons interacting in the upper half of the SDD will be recorded by the upper row of anodes, and X-ray photons interacting in the lower half will be recorded via the lower row of anodes. Although there are only four SDDs, the Back-End- Electronics must be able to handle signals from 8 separate detector halves.

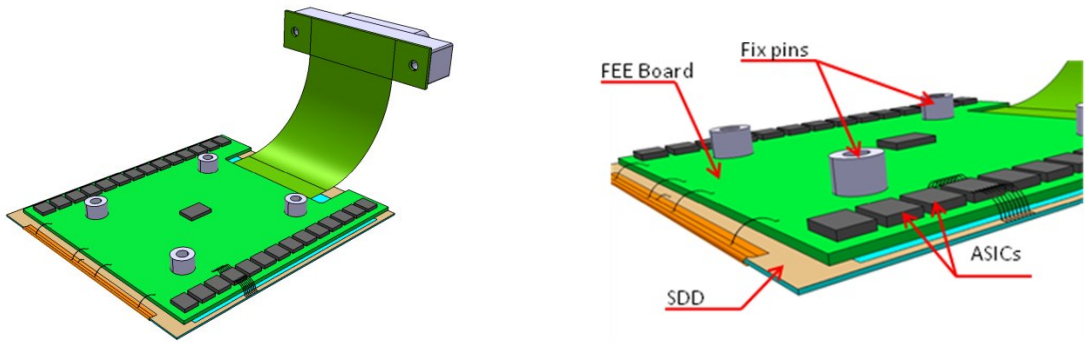

Figure 11 A view of the front end electronics (FEE) board and the ASICs mounted on the back side of one of the four SDD detector modules in one WFM camera

### 5.3 Back End Electronics

The WFM Back-End-Electronics differs significantly from the corresponding Module-Back-End-Electronic of the LAD [20] because the WFM needs to calculate the position of each photon in the detector plane accurately. The WFM-BEE main functions are described in section 0 and illustrated and illustrated in Figure 3. The WFM-BEE collects the HK data from the sensors placed inside the camera housing and controls the detector power supply board and heaters.

The BEE board is based on a VIRTEX 4 FPGA (XQR4VSX55) device driven by a 60 MHz system clock. The chip is radiation tolerant with TID >300 kRad(Si) and Dose Rate Latch-up >10$^7$ Rad(Si)/s. All functionalities will be implemented in the FPGA firmware.

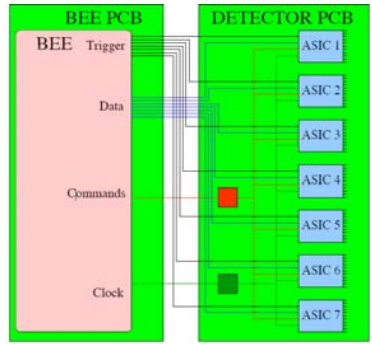

Figure 12 Interface between one half SDD/FEE and the BEE.

## 5.4 Instrument Control Unit

The WFM instrument control unit (ICU) will contain the functionalities of instrument control and data handling and include the mass memory, as well as the power distribution unit (see Figure 3). The central data handling unit will be based on the Virtex-5QV (XQR5VFX130) FPGA device driven by a 450 MHz system clock and provides >100 DMIPS when a modular, ESA-Qualified LEON3FT VHDL IP Core is implemented. The architecture of the WFM data handling unit is outlined in Figure 13. The chip has ample radiation tolerance: TID >1Mrad (Si) and Dose Rate Latch-up >$10^{10}$ Rad(Si)/s. Two identical ICUs are provided with one in cold redundancy.

The LOFT burst onboard trigger (LBOT) functionality [16] will be implemented in the FPGA in a dedicated 2D FFT processor for coded mask image deconvolution based on standard IP core modules. This is thus an integral part of the DHU. Current estimates of the FPGA resource utilization indicate that less than 50% of the available logic units are used (including margin). Although not foreseen to be needed, this would allow for a dual core LEON3FT to be implemented.

The memory allocation includes 1MB non-volatile memory in EEPROM for storage of flight software and 4 MB System Memory in SRAM. Mass Storage on the DHU board is implemented in 133 MHz SDRAM with EDAC organized in 48-bit words, having a capacity of 24 Gbits (2GB data). The FPGA configuration image resides in a 64 Mbit (8MB) PROM. An additional 1MB of system memory may be implemented directly on the FPGA using Block RAM modules.

The WFM power distribution unit (PDU) provides switched, protected main bus power lines to the BEE units in each of the 10 WFM cameras. The PDU latching current limiters (LCL) include ON/OFF-switching, current limitation and automatic switch-off if an over-current condition persists after a fixed delay (e.g. 10 ms). Under-/over-voltage protection is also included in the LCL functionality.

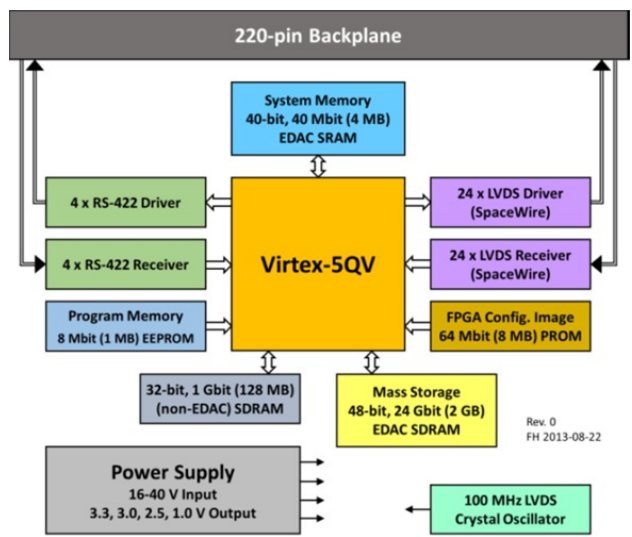

Figure 13 The WFM data handling unit architecture centered on the Virtex-5 FPGA

# 6. WFM MECHANICAL AND THERMAL DESIGN

The mechanical design of the WFM camera is driven by the need to establish and maintain flat and parallel the plane of the coded mask and the plane of the detectors. This is essential to achieve stable and accurate imaging properties of the system. The thermal design of the WFM is also driven by this concern, as well as the need to maintain an operational temperature of the SDD detectors in the ÷30°C to ÷3°C range. The typical flatness and tolerances are of the order of ±50 µm. An overview of the main camera components is shown in Figure 14.

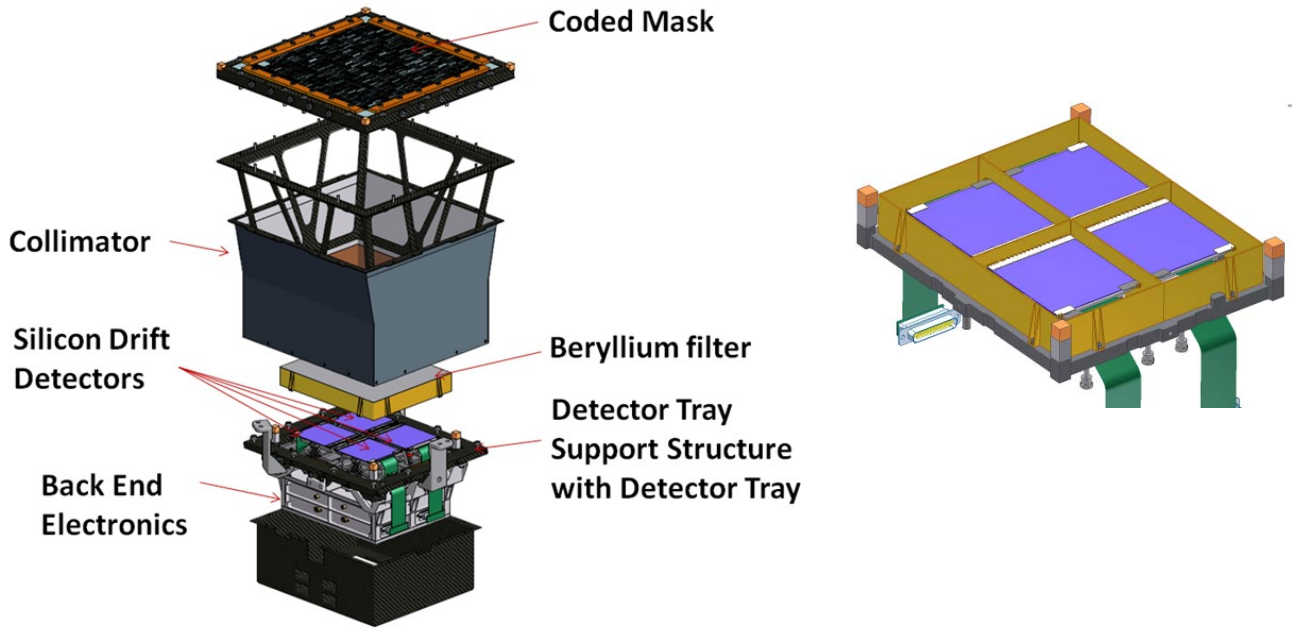

Figure 14 Left: an exploded view of a single camera with the main mechanical sub-systems. Right: the 4 SDD detector modules forming the detection plane.

## 6.1 Coded Mask and Frame

The 150 µm thick Tungsten coded mask (described in section 4.3) is mounted in a frame. In order to minimize the vertical displacements of the mask during operations a pretension mechanism is introduced. The mask frame consists of CFRP and fiberglass, where the fiberglass material has been chosen because it has a CTE similar to that of the Tungsten mask.

## 6.2 Collimator

The collimator is the mechanical link between the detector plane and the mask, and serves to block X-rays coming from outside the desired FoV. The collimator supports the coded mask frame assembly and is made of 3 mm open CFRP structure (see Figure 14). This structure of the collimator permits to have access to the detector tray after the complete assembly of the camera and has enough stiffness to avoid deformations that can appear during launch (accelerations) and operation of the WFM (thermal stresses). The collimator will be covered by a shield: 1mm thick CFRP structure covered from the outside by Tungsten (150 µm) plus Copper (50 µm) plates and Molybdenum (50 µm) plates from the inside. The Copper and Molybdenum foils are included for in-flight calibration purposes (see section 8.2).

## 6.3 Detector Tray and Support Structure

The detector tray assembly consists of 4 detector module assemblies and the detector support plate (see Figure 14 and Figure 15). In the detector assembly the SDD tile is glued to a ceramic PCB containing the front-end electronics (FEE). For an optimal compact design the FEE will mainly consist of a ceramic substrate with roughly the same dimensions as the detector tile, directly glued onto the bottom of the detector into a sandwich. The ASICs reading out the charge from the anodes of the detector are situated on the two edges of the FEE, as close as possible to the anodes of the SDD for a minimal wire length (see Figure 11). Crucial for alignment stability is the thermo-mechanical behavior of assemblies over temperature. For this reason Aluminum Nitride has been selected as substrate for the FEE. The coefficient of thermal expansion (CTE) of Aluminum Nitride is 4.6ppm/K, which is relatively close to that of Silicon (2.6ppm/K). The detector module sandwich is supported by an invar bracket with an Al plate in between connected to a cooling strap for thermal control of the detector sandwich.

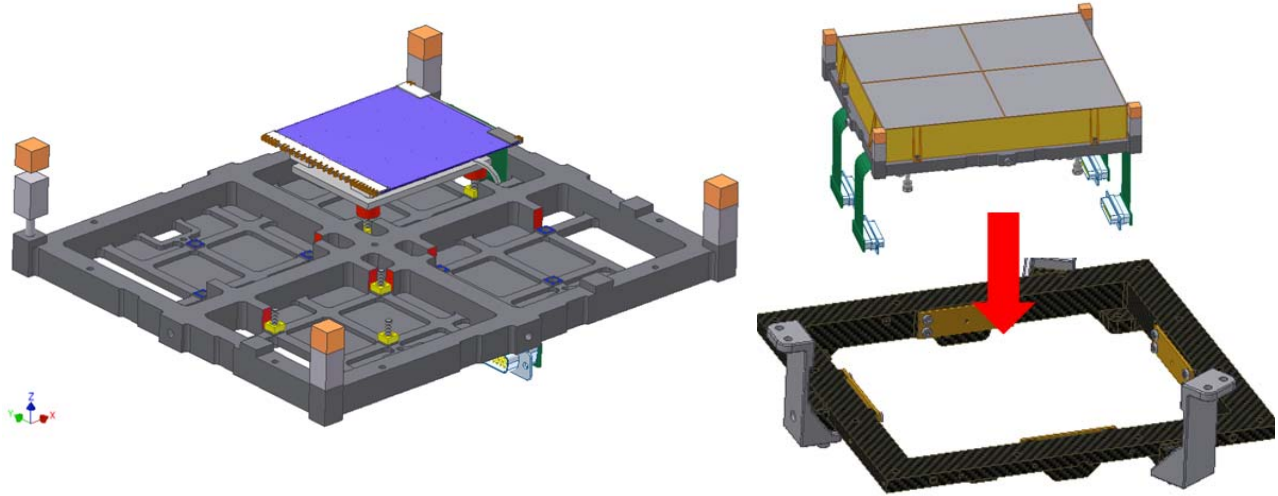

Figure 15 The detector support plate with one detector module (left). The 4 optical alignment cubes are shown in orange. The detector tray assembly with 4 detector modules is integrated into the detector tray support structure (right)

The detector support plate containing the 4 SDD detector assemblies (see Figure 15, left) is also made of Invar in order to match the CTE of the detector assembly. Each of the four detector assemblies features 6 alignment reference surfaces, which will be used for mechanical alignment. For the alignment of the detector plane with the coded mask frame four optical cubes are mounted on the detector support plate (see Figure 15).

The detector tray support structure (see right part of Figure 15) serves as a support for the detector support plate, and facilitates the mounting of the collimator and the back end electronics box. Also the iso-static fittings are mounted on the detector support structure and serve as mechanical interface of the WFM Camera to the spacecraft.

The support structure is made of CFRP in order to avoid displacements due to on-orbit thermal variations, which affect the alignment of the camera. The placement of the detector support plate inside the detector support structure will be carried out using four leaf springs (the so called "over-constrained" iso-static mounting).

### 6.4 Berylium windows

Studies have shown that the impacts of micro-meteorites and small particles of orbital debris will pose a risk for the Si detectors, due to the large FoV of the WFM camera. To mitigate this effect, a thin (25 μm) Beryllium filter will be placed ~8 mm above the detector plane. In combination with the thermal blanket in front of the coded mask, this Be filter will reduce the risk of impact of particles of size ~100 μm to about $1.6\times10^{-2}$ per year per camera. The low energy response is only mildly affected by the Be filter, which has a transparency of ~70% at 2 keV. See [12] for detailed studies of risks posed to LOFT by micro-meteorites and orbital debris.

### 6.5 Back End Electronics box

The back end electronics box is integrated at the bottom of the camera, where the power supply board is placed as well (see Figure 14). The BEE box is made of aluminum. A light tight cover made of CFRP will cover the lateral part of the BEE box, where the electrical connectors for the harness to the central instrument control unit are also placed.

### 6.6 WFM Thermal Design

The WFM camera contains 3 separate elements requiring independent thermal control designs. These elements are: the mask, the 4 independent SDD/FEE sandwiches, and the BEE box located at the bottom of the camera. A schematic of the heat transfer processes is shown in Figure 16.

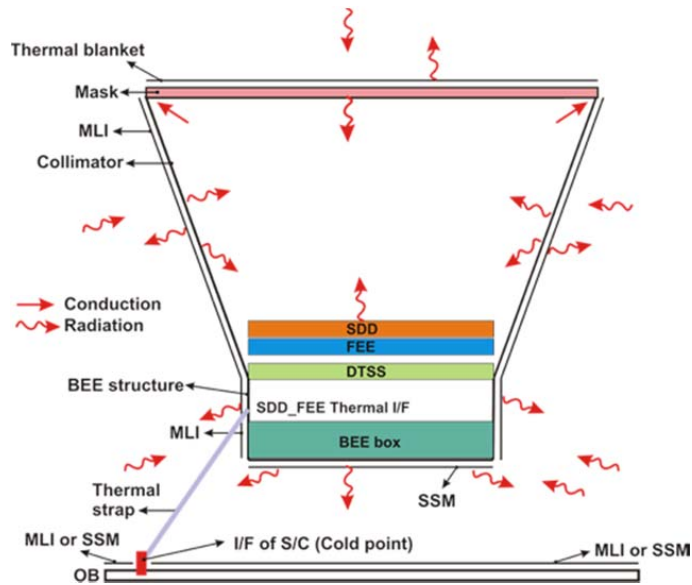

Figure 16 WFM camera heat transfer processes scheme with the thermal strap connected to the spacecraft structure

The mask is covered by a Kapton thermal foil (Sheldahl 146545, 0.15μm $SiO_2$ + 100nm VDA + 7.6μm Kapton), which has an acceptable transparency to X-rays at 2 keV. The heat dissipation of the detector plane will mainly be achieved by conduction through a thermal strap provided by the spacecraft. The Back End Electronics box will mainly dissipate its heat from the back side of the box, which acts as a radiator.

A MLI blanket will wrap the camera in order to provide an optimum isolation to the SDD/FEE sandwich and to keep the camera alignment in the desired margin during the mission The source positioning accuracy of the camera will depend on thermal stability of the structure and therefore a Sun shade will be provided by the spacecraft to prevent that direct light from the Sun directly illuminate the WFM camera assembly.

## 7. WFM DATA AND TELEMETRY

The WFM overall data flow is illustrated by the functional diagram in Figure 3. The signals from the ASICs in each of the 10 cameras are collected and processed in the associated back end electronics (BEE). The main task of the BEE is to process the anode signals when the ASIC is triggered by an interaction in the detector. Unwanted triggers from particle interactions are filtered out. By fitting the Gaussian shape of the charge cloud signals after proper pedestal and common noise subtraction, the position and energy of the incoming photon is determined. The X-ray event data (X,Y,E,T) also includes the time of the interaction with an accuracy of ~5μs. This event data packet can be transmitted in 40 bits per event by applying differential time tagging.

### 7.1 Normal data taking

The maximum, sustained data rate for the WFM will normally be limited by the down-link bandwidth to ~100 kbits/s averaged over the orbit. Therefore it will not be possible to downlink the full event-by-event information and some data binning is normally required. The normal standard data mode will consist of 3 binned data products:

- Detector images integrated over 300s in 8 or 16 energy bands;
- Full resolution detector spectra integrated over 30s;
- Detector rate-meter data with 16 ms resolution in 8 energy bands.

The bulk of these data consists of the detector images. The images are for typical data rates and more than 8 energy bands sparsely filled (much less than 1 count per pixel). This allows for efficient compression by transmitting the distance between filled pixels. The optimum number of bits per event used in the encoding is easily determined by the count rate. The highest rates are expected from the camera pairs observing the Galactic Center region with several bright sources. Depending on the final observing strategies, the parameters for image integration time and/or the number of energy bands may be adjusted dynamically or by command in order to optimize the telemetry usage. We notice that the difference between the telemetry rates for 8 and 16 energy bands is not large. This is explained by the simple fact that the distance between filled pixels will double when switching from 8 to 16 energy bands, thus requiring only 1 more bit per event to encode the distance between filled pixels.

### 7.2 Event-by-event data and trigger mode

X-ray events of duration significantly shorter than the normal integration time for detector images of 300 s will be difficult to identify in the normal data. Therefore, a burst trigger logic will operate to detect increases in flux on time scales from a fraction of a second to ~100s. The burst trigger logic is active during data taking mode and will be able to distinguish between increases in the count rate of X-ray events in each WFM cameras due to celestial sources and background. When the onboard software detects a potential transient event it will save the data for transmission in the "event-by-event" format in order for the ground software to be able to analyze the event in full detail. A burst trigger will mark a 300s-long interval of data around the trigger time to be transmitted in event-by-event format.

### 7.3 The LOFT Burst Alert System

It is expected that the WFM will detect ~120 Gamma Ray Bursts per year. Scientifically it is highly desirable to observe these sources with other telescopes and instruments as soon as possible after (or even during) the event. Therefore LOFT will employ a VHF transmission capability to send a short message about the occurrence of such events with minimum delay to a network of VHF receiving stations on the ground for further distribution to interested observatories. The LOFT burst alert system will distribute the location of a transient event with ~1 arc minute accuracy to end users within 30 s (goal 20s) of the onboard event detection in at least 2/3 of the cases.

The onboard software will localize the position in the sky of the source responsible for the burst trigger. For a coded mask instrument, the deconvolution of the detector image into a sky image is computationally very demanding. The deconvolution will be done onboard by cross correlation of the mask pattern with the background subtracted detector image. This is most efficiently done by a discrete Fast Fourier Transform (DFFT) algorithm. As a result of a burst trigger, the onboard software will enter a mode to determine the position of the source. The DFFT transformation from coded mask detector images into sky images will be performed the data handling unit FPGA optimized to perform the DFFT. The burst alert system is described in detail in [16].

### 7.4 The WFM on ground data processing

The LOFT ground segment will include the science data center (SDC), which is described in detail in [1]. For the SDC a high priority task will be to analyze the WFM data immediately, or as soon as possible, after the reception of the science data from the mission operations center (MOC) in order to identify new sources and state changes in known objects, that will warrant a target of opportunity for the LAD. The quick-look analysis results obtained from the WFM data are also expected to serve as input for target-of-opportunity for other space or ground based observatories.

## 8. WFM CALIBRATION

### 8.1 On ground characterization and calibration

The main characterization of the WFM detectors will be performed when the 4 SDD modules are integrated to form the camera detector plane, but before the mounting of the collimator and the coded mask. The parameters to be characterized are listed in Table 5.

The assembled WFM cameras will be characterized with respect to their imaging performance, determining the camera PSF and angular response. The calibration campaign will also determine the WFM camera sensitivity as well as the so called coding power or coding efficiency of the coded aperture cameras. The coding efficiency is a measurement of the perfectness of a coded mask and detector system. The coding efficiency for an ideal system is unity, but decreases to a value below 1 due to mask and detector misalignments, tilts and rotations between mask and detector plane.

Table 5 WFM detector parameters to be characterized and calibrated on detector plane level

| ITEM | Parameters to be calibrated/characterized | Environmental Conditions | Calibration Accuracy |
|---|---|---|---|
| **Energy scale calibration** | Offset and gain of each SDD/ASIC channel as a function of temperature. | ÷50° C < T < 0 °C (10 °C step) in vacuum | <1% |
| **Energy range verification** | Lower energy threshold of each SDD/ASIC channel for one representative temperature. | One representative temperature in vacuum | |
| **Spatial resolution calibration** | Position resolution as a function of photon energy and absorption point. | ÷50° C < T < 0 °C (10 °C step) in vacuum | < 10% |
| **Energy resolution characterization** | Width of monochromatic line spectra as a function of photon energy and absorption point | ÷50° C< T < 0 °C (10 °C step) in vacuum | 100 eV |
| **Spatial resolution uniformity** | Verification that the capability to reconstruct the absorption point of the photon is uniform | One representative temperature in vacuum | |

Performance tests and calibration of the WFM coded mask cameras would ideally require non-divergent X-ray beams. The effect of beam divergence will be handled with the techniques developed in connection with the SuperAGILE calibration campaign, see [6].

## 8.2 In orbit calibration

The in-orbit energy calibration will confirm/monitor the overall energy response of the WFM cameras during the mission. This will be achieved by electronic stimulation of the WFM ASICs to monitor the electronic response (off-set and gain) of each ASIC channel and by detection of weak X-ray lines. The purpose is not to calibrate each individual detector channel, but rather monitor the overall energy calibration for each WFM camera/SDD.

The X-ray illumination of the detector planes could be made with the help of radioactive sources or an X-ray generator. These methods have various drawbacks, and studies have showed that using fluorescence lines produced by the LOFT orbit ambient radiation (mainly the CXB) is a viable alternative. The main part of the fluorescence radiation reaching the detector is generated by the CXB radiation which interacts with material inside the collimator. A design where the inside of the collimator is covered with 50 μm Cu and Mo foils to generate suitable calibration lines was found adequate. A schematic of the camera used in a GEANT-4 simulation and the resulting background spectrum is shown in Figure 17. The general model used for background simulations for both the LAD and WFM is described in [1] [3]. A simulation of a camera background spectrum with 2 minutes of integration time is shown in Figure 17. Here the Cu K-shell lines at 8.05 and 8.90 keV, as well as the Mo K-shell lines at 17.4, 17.5 and 19.6 keV can be clearly seen. The WFM camera does not need onboard radioactive sources for in orbit calibration.

The WFM imaging performance will be calibrated through observations of celestial X-ray sources. These will allow determination of the WFM camera PSFs, the WFM camera pointing (bore-sight) and yield the source location accuracy of the telescope. Most cataloged X-ray sources have positions known to better than 1 arc sec. The statistical position errors for many sources will be better than a few arc sec in the central FoV and better than 10 arc sec in the full FoV (in 1-day observation). Many sources will also be viewed simultaneously by more than 1 camera pair in the overlapping fields of view.

The observations of these bright sources with known positions will provide in orbit determination of the bore sight and image parameters with accuracy in the range of a few arc sec within the first few months of operations. The statistical accuracy will improve over time and each WFM observation of sources with known positions will contribute to the improvement and verification of the position calibration. Any dependence on parameters like temperatures and orbital phase can also be mapped and subsequently corrected for. The overlapping fields of view between different camera pairs will also provide a good handle of the in-flight calibration of the off-axis response, as sources will be simultaneously observed at different off-axis angles in individual cameras.

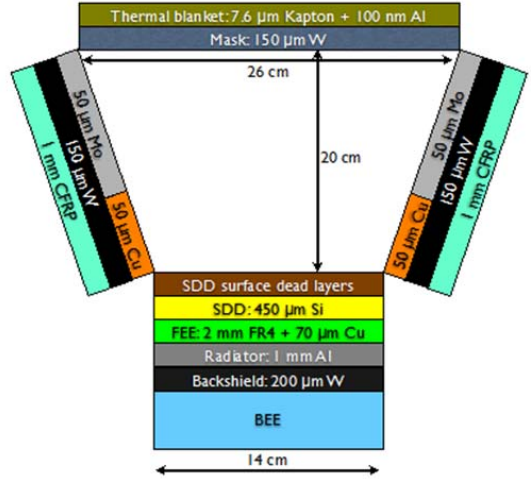 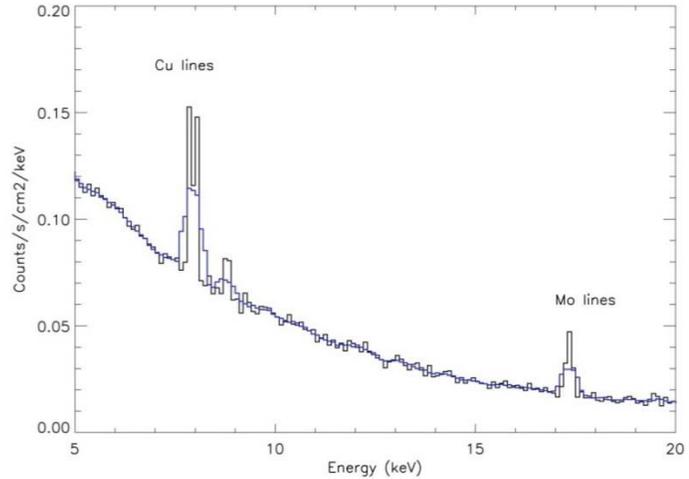

Figure 17 *Left:* the geometry of the WFM camera used to simulate the fluorescence lines useful for in-orbit calibration by introducing 50 μm layers of Mo and Cu on the inside of the collimator (left). *Right:* the resulting background spectrum with lines suitable for in-orbit calibration (right). An integration time of 2 minutes is assumed and a detector resolution of 0.5 keV has been applied for the blue curve.

## 9. CONCLUSIONS

Although not selected for the ESA M3 mission, the LOFT mission concept promises to significantly advance our knowledge about some of the fundamental questions related to strong gravity and matter at supra-nuclear densities. The LOFT consortium is thus planning to continue seeking opportunities to continue developing this mission. The ESA M4 competition is being evaluated, together with alternative options.

We have presented here the consolidated design of the WFM instrument achieved during the LOFT assessment study phase. At present, the WFM configuration has been tuned to fulfil the LOFT science goals. However, the high modularity of the WFM design makes this instrument particularly well suited to be used on any future mission that aims at making use of wide field monitoring capabilities to enhance the return of narrow field instruments. Although designed as a support instrument, the WFM is expected to provide important independent contributions to the study of the gamma ray burst through the near real time burst alert system.

## AKNOWLEDGEMENTS

IEEC-CSIC has been supported by the Spanish MINECO (under grant AYA2011-24704). MSSL-UCL is supported by the UK Space Agency. SRON is funded by the Dutch national science foundation (NWO). The group at the University of Geneva is supported by the Swiss Space Office. The Italian team is grateful for support by ASI (under contract I/021/12/0-186/12), INAF and INFN. The work of IAAT on LOFT is supported by Germany's national research center for aeronautics and space (DRL). The work of the IRAP group is supported by the French Space Agency. ECAP is supported by DLR (under grant number 50 OO 1111).